# Simulations and measurements of the impact of collective effects on dynamic aperture.

V. Sajaev,[1] R. Lindberg,[1] S. Shin,[2,*] and M. Borland [1]

[1] Advanced Photon Source, Argonne National Laboratory, Illinois 60439 USA
[2] Pohang Accelerator Laboratory, POSTECH, Pohang, Kyungbuk, 790-784, KOREA

E-mail: tlssh@postech.ac.kr

**Abstract**. We describe a benchmark study of collective and nonlinear dynamics in an APS storage ring. A 1-mm long bunch was assumed in the calculation of wakefield and element by element particle tracking with distributed wakefield component along the ring was performed in Elegant simulation. The result of Elegant simulation differed by less than 5 % from experimental measurement.

## 1. Introduction

The ultimate storage ring (USR) based on the multi-bend achromat (MBA) lattice concept is emerging to push beyond the brightness and coherence reached by present third-generation storage ring. MAX-IV [1] was the first MBA machine; its successor, Sirius [2], is currently under construction. Other projects are being conducted to convert existing third generation machines such as ESRF-II, APSU, Spring-8-II and ALSU [3, 4, 5, 6] to USRs. In these facilities, beam emittance goes down to a few hundred picometers, or even ≤ 100 pm.

When the emittance of storage ring is pushed to an extremely small value (i.e., ≤ ~ 100 pm), two primary consequences emerge: the dynamic aperture becomes insufficient to enable accumulation based injection, so only on-axis swap-out injection [7] is workable; and the maximum single bunch intensity is significantly reduced due to the intra-beam scattering (IBS) and beam instabilities. Recent simulations [8] indicate that collective effects are critical issue on requirement to achieve beam injection with high charge. Collective effects may preclude accumulation of 4.2 mA/bunch in a ring without transverse bunch by bunch feedback system in APSU. These are critical issues for operation, so careful modeling, simulation and benchmark study must be conducted to quantify them.

A precise model of an accelerator ring is crucial to achieve ultimate performance in a storage ring. LOCO algorithms [9] have been developed to calibrate the linear model, and have been successfully applied experimentally to determine and correct the linear optics of the ring. Based on the success of this process of modeling linear optics, storage-ring models that consider non-linear optics and collective effect with impedance have been proposed, and used to build a realistic model of the storage ring. Frequency analysis of betatron motion [10] is used to extract information about the resonance driving term that affects the non-linear dynamics of the particle beam. Multi-particle tracking simulations [11] with the model including wakefields have also been performed to predict intensity-dependent beam characteristics in a storage ring. Unfortunately, some results on collective effect in many storage rings are significantly different between model predictions and experimental measurements. Furthermore, frequency analysis of betatron motion is seriously limited by noise from the beam position monitor (BPM) and by decoherence of the excited betatron oscillation. Therefore, accurate prediction of the dynamics of a high-charge injected beam in a USR is an extremely challenging task.

Impedance databases have been thoroughly updated, and models of impedance have been improved in APS. In addition, an accelerator code, Elegant [12], has been continuously upgraded to model USR by using many beam dynamics functions (e.g., intra-beam scattering, bunch-by-bunch feedback, ion gap effect). Based on this background, many benchmark studies for APS-U have been performed at APS [13]. One important benchmark study considers a kick aperture to demonstrate the dynamics of a high-charge injected beam in APSU. A benchmark study of collective and nonlinear dynamics in APS storage ring using kicked beam was performed successfully using Elegant and the APS impedance database. The result of Elegant simulation differed by < 5% from experimental measurement.

In this paper, we describe a benchmark study and its result for simulation of a high-charge injected beam in APSU. Section 2 describes how to perform impedance modeling in APS, then introduces a method to track particles with collective effect. Section 3 describes Elegant simulation and its result for the kick aperture. Section 4 gives the comparison between measurement and simulation, with a beam dynamic explanation. Finally, Section 5 presents conclusions.

## 2. Impedance model in APS

The effects of impedances and wakefields are modelled using a single impedance element or using several elements in Elegant. To reduce the distributed impedance from the entire ring to a localized perturbation, the impedance is first divided into its resistive wall and geometric components. The resistive wall contribution is calculated by using analytic formulae applied to the round chambers. In contrast, because the geometric impedance depends on the detailed cross-sectional variation of various components along the vacuum vessel, the computation of geometric impedance requires additional numerical simulation tools and methods.

The geometric contribution is built from the individual impedances and wakefields generated by various elements around the ring. The wakefields that are produced by axially-symmetric components are computed using 2-D ECHO code [14], whereas the more-general wakefields from structures that vary in 3-D are calculated using the finite-difference time-domain code GdfidL [15]. To balance numerical efficiency and accuracy, each of these codes models the point-particle wakefields by using the wake potential generated by a 1-mm long bunch, because this approximation has had success in predicting the onset of various instabilities in the current APS. In addition, numerical tests have been performed by using wake potentials derived from electron bunches < 1 mm long, and these have proven to give the same single-bunch current limit. The total transverse wake potential of the ring is found by weighting the individual geometric contributions by their respective local lattice functions, then summing. More precisely, if each element is represented by j and the lattice functions at that element are represented by $\beta_{x,j}$ and $\beta_{y,j}$, the weighted geometric wakefields along x and y directions are given by

$$<\beta_x W_x^{geo}> = \sum_{elements\ j} \beta_{x,j} W_{x,j}^{geo} \qquad <\beta_y W_y^{geo}> = \sum_{elements\ j} \beta_{y,j} W_{y,j}^{geo} \qquad (1)$$

The total longitudinal wakefield is the simple sum

$$<W_z^{geo}> = \sum_{elements\ j} \beta_{z,j} W_{z,j}^{geo} \qquad (2)$$

The corresponding impedances are then computed using the discrete Fourier transform. Finally, the total impedance is obtained by adding the resistive wall contribution to the geometric impedance, then applied in the particle tracking code Elegant as a single impedance element or as several impedance elements by dividing by the lattice function at its chosen locations.

Longitudinal wakefield change the test particle (particle behind drive electron) energy

$$\Delta \gamma_{test} \approx -\frac{e}{mc^2} W_z(z_{drive} - z_{test}) \qquad (3)$$

The transverse wakefield change the test particle angle

$$\Delta x'_{test} \approx -\frac{e}{\gamma mc^2}[W_{x,M}(z_{drive} - z_{test}) + x_{drive} W_{x,D}(z_{drive} - z_{test}) + x_{test} W_{x,Q}(z_{drive} - z_{test})] \qquad (4)$$

Here the first term is a monopole wakefield that is generated by a chamber that is not mirror symmetric; this term and can cause emittance to increase. The second term is dipole wakefield that is a source of collective instabilities scales with displacement of drive electron. The third term is the source of mainly tune-shift scales with displacement of test electron. The sum of the longitudinal and transverse impedances in the APS storage ring (Figure 1) was both inductive and resistive from low frequencies to high frequencies. The impedance distributions in the two transverse planes were similar in APS.

## 3. Elegant simulation for kick aperture

A method to calculate impedances and to use these results in predictive simulations of the storage ring has been developed and implemented at APS. This impedance model has successfully reproduced various impedance-driven collective effects observed in the APS ring. To demonstrate simulation of the injected beam in APSU, we extend the previous scheme to simulate kick aperture at various charges. In this section, we describe the simulation method and result for the kick aperture.

Elegant is very powerful tool to simulate an injection beam with high charge for USR, and is used to simulate a storage ring that uses a kicked beam. For optics calibration, the values of quadrupole and sextupole obtained during machine study are implemented into Elegant. Here quadrupole set and tilt were obtained to reproduce quadrupole error and coupling error in real machine. For particle tracking, all elements in the APS storage ring are included in the model. Classical synchrotron radiation is calculated in all magnets; the one-turn radiation energy loss by a dipole is 5.353 MeV. Energy gain from the RF cavity is also included. For simplicity, only one RF cavity in Sector 40 represents all APS RF cavities distributed in sectors 36, 37, 38, and 40. The RF frequency is 351,933,984 Hz, the harmonic number is h = 1296 and total RF voltage is 9.5 MV.

The changes in the closed orbit as consequences of finite energy gain in the RF cavity in the dispersive region, and of radiation energy losses can be described by a Green's function [16]

$$x_{c.o,d}(s) = \sum_i G(s,s') \frac{\Delta p(s_i)}{p_0} = -\frac{\sqrt{\beta(s)}}{2\sin\pi\nu_x} \sqrt{H(s')} \cos\Psi(s,s') \frac{\Delta p(s_i)}{p_0}, \tag{5}$$

where

$$H(s') = \frac{1}{\beta(s')} \{\eta(s')^2 + [\alpha(s')\eta(s') + \beta(s')\eta'(s')]^2\} \tag{6}$$

$$\Psi(s,s_i) = \psi(s) - \psi(s_i) + \chi(s_i), \quad \chi(s_i) = \tan^{-1} \frac{\eta(s_i)}{\alpha(s_i)\eta(s_i) + \beta(s_i)\eta'(s_i)} \tag{7}$$

For each particle with an arbitrary energy deviation there exists a closed orbit that is given by the product of the dispersion function times the relative momentum deviation [17]. Therefore, if finite energy gain (through RF cavity) and loss (through bending magnet) in dispersive region and momentum deviation through element by element are considered, total closed orbit is given by

$$x_{c.o}(s) = x_{c.o,d}(s) + \eta_x(s) \frac{dp}{p}(s) \tag{8}$$

In Eq. (5) and (8), the effect by kicks through bending magnets average out since bending magnets are widely distributed along phase advance in Eq. (7). But the effect through RF cavity is significant due to the localized kick. Momentum deviation along the ring is shown in Fig. 2. For simplicity, one RF cavity at the end location of the ring compensate total energy loss through all bending magnets. This momentum deviation is also periodic along the turn by turn. Therefore, closed orbit distortion is resulted by the kick from RF cavity and the change of momentum deviation along the ring. Figure 3 shows closed orbit distortion by kick from RF cavity and the change of momentum deviation. There is a coupling from the betatron oscillation to the synchrotron oscillation which is caused by the path lengthening due to the betatron oscillation. Therefore, a few μm beam oscillation around closed orbit of Fig. 3 is also occurred in the simulation.

In order to generate equilibrium particle distribution, pre-particle tracking is performed up to 10,000 turns with 10,000 particles generated from the lattice information. After that, equilibrium particle distributions as initial beam are used in the simulations for kick aperture. Generally it is enough to consider one impedance element along the ring for simulating wakefield effect on beam since synchrotron frequency is much longer than revolution frequency. But if betatron tune is getting closer on integer, results between one impedance element and distributed impedance elements along the ring are obviously different [18]. Therefore, we divided one impedance element to 40 distributed impedance elements in every straight section. Measured kicker profile is modelled in the simulation. It has a 3.6 µs half-sine wave so that the single bunch received only single kick from kicker. In order to investigate the impact of collective effects on dynamic aperture, beam loss by exciting the beam to large amplitudes with kicks are simulated along different single bunch charge and results will be described and compared with experiment.

## 4. Kick aperture measurement and analysis

Data for the kick aperture were collected on January 2017 and beam parameters for experiment are introduced in Table 1. First, linear optics was calibrated with the model and confirmed after calibration. The stable range of a single-bunch current was explored at set chromaticity $\xi_x/\xi_y$ = 6.6/5.9; a single bunch was stable up to 5.33 mA. Kick apertures were measured with bunch currents of 0.9, 2.3 and 4.1, which are all < 5.33 mA, so equilibrium particle distributions (Section 3) can be used in the simulations.

First, betatron oscillations that a kick imparts to a single bunch were measured using a turn-by-turn beam-position monitor. Comparison of betatron oscillation after the kick, as obtained by measurement and simulation (Fig. 4) shows that decoherence was significant in the horizontal plane, and that the amplitude of beam oscillation due to coupling is > 0.5 mm in vertical plane. In each plane, the tunes in the simulation were corrected to have the same phase advances as detected in the measurement; all other parameters in the simulations were the same as used during the measurements. The decoherences in measurement agree well with those in simulation.

Kicker strength was calibrated with beam response by kick in simulation and measurement. To avoid the decoherence effect in the calibration, only the first-turn data after the kick (Fig. 5a) were considered. Data from all BPMs and kicks of 2, 2.5 and 3 kV (Fig. 5b) were used to filter certain possible errors along the measurement axis and the simulation axis. In simulations, 1 kV was assumed to correspond to a 0.1-mrad kick. By linear fitting of data in Fig. 5b, the calibration factor was estimated to be 1 kV = 0.111 mrad.

The variation of tune with the action of single particle (i.e., Amplitude detuning) can be described by a Taylor expansion as

$$v_z(J_x, J_y) = v_{z0} + \frac{\partial v_z}{\partial J_x} J_x + \frac{\partial v_z}{\partial J_y} J_y + \frac{1}{2}(\frac{\partial^2 v_z}{\partial J_x^2} J_x^2 + 2\frac{\partial^2 v_z}{\partial J_x \partial J_y} J_x J_y + \frac{\partial^2 v_z}{\partial J_y^2} J_y^2) + \cdots \quad (9)$$

If small action in the vertical plane is considered due to horizontal kick, amplitude detuning is given by

$$v_x(J_x) = v_{x0} + \frac{\partial v_x}{\partial J_x} J_x + \frac{1}{2}\frac{\partial^2 v_x}{\partial J_x^2} J_x^2 + \cdots \quad (10)$$

To quantify the amplitude detuning, measured and simulated beam oscillations data turn-by-turn were used for every kick. Due to large decoherence, FFT of beam oscillation does not estimate tune well, so tunes were determined from zero-crossing points in Fig. 4. Horizontal detuning varied with amplitude (Figure 6). Five turn data of beam oscillation after the kick were interpolated to estimate zero-crossing points, and four tunes were determined from point-to-point phase advances. Parameters of a second-order polynomial fit to the curve in Fig. 6 agree reasonably well with simulation data (Table 2).

The dynamic aperture is the boundary in phase space beyond which particle motion becomes unstable. This aperture can be measured from the beam loss following large-amplitude kicks. To measure dynamic aperture of the APS storage ring, we used a fast horizontal kicker. Kicker voltage was scanned and the fraction of the beam that survived the kick was recorded based on current measurements. The measurement was performed with a single bunch in which coupling was with 1 %. Turn-by-turn BPMs are used to measure kick amplitude. The impact of collective effects on dynamic aperture was measured from beam loss following large-amplitude kicks for three different initial single-bunch currents. In measurements, the amplitudes of 50 % beam loss for three different single bunch currents were almost the same (~ 0.44 mrad). A single bunch with higher initial current has more beam loss in the kick region

< 0.44 mrad and has less beam loss in the kick region above > 0.44 mrad. Simulation and measurement results agreed well (Fig. 7).

Understanding of beam loss tendency along the beam current and kick would provide useful insights. Figure 8 shows schematics to describe beam loss phenomenon. Dynamic aperture, which is stable area of beam motion in phase space, is represented as black line in the figure. A large kick takes the beam to large amplitude in the momentum axis of phase space. If the kick is large, the beam passes beyond the dynamic aperture range, and is lost; if the beam is at the boundary of dynamic aperture, some of the beam can be lost. Figure 8 (a) and (b) shows kicked beam below and above kick aperture, respectively. Higher beam current have large emittance and vice versa in both cases. Although unfortunately it is impossible to describe emittance increase along charge exactly, normal mode analysis for two particle model was used to describe the schematics in Fig. 8 [19].

$$\begin{pmatrix} \hat{y}_1(s = 2nc\pi/\omega_s) \\ \hat{y}_2(s = 2nc\pi/\omega_s) \end{pmatrix} = e^{n(iRe\{T\}-Im\{T\})} \left(\frac{\hat{y}_1(0)+\hat{y}_2(0)}{2}\right)\begin{pmatrix} +1 \\ 1 \end{pmatrix} + e^{n(-iRe\{T\}+Im\{T\})}\left(\frac{-\hat{y}_1(0)+\hat{y}_2(0)}{2}\right)\begin{pmatrix} -1 \\ 1 \end{pmatrix} \quad (11)$$

where $\hat{y}_1(s)$ is a slow varying complex function of $n$ and

$$Im\{T\} = \frac{Nr_0Wc}{4\gamma C\omega_s}\frac{\xi}{\eta} \quad (12)$$

The term $e^{n(iRe\{T\})}$ implies just oscillation along the turn. When $Im\{T\} > 0$ (positive chromaticity), the term $e^{n(Im\{T\})}$ increases exponentially in $\begin{pmatrix} -1 \\ 1 \end{pmatrix}$ mode (head-tail mode) and decrease exponentially in $\begin{pmatrix} +1 \\ 1 \end{pmatrix}$ mode (centroid mode). Therefore, when chromaticity is high and positive, projected beam emittance increases and centroid motion is almost the same along the beam current, as in Fig. 8.

In real situations, the wakefield effect causes motion of the kicked beam to be so complicated that it should be described numerically rather than analytically. Figure 9 shows emittance increase and particle distribution in phase space along turn. This result justifies the schematic descriptions in Fig. 8. Emittance increases as charge increased below and above aperture (Fig. 9). Analytically, when chromaticity is positive, head-tail motion is dominant and increases projected emittance. Beam distribution in phase space spreads out as a consequence of turn spread at high chromaticity, so emittance of a low beam current also increases along the turn after a kick. Particle distribution in phase space was observed during 10 turns, with the kicker fired between the second and third turns; the result (Fig. 9) describes the process of decoherence and emittance increase after the kick. The figure shows an island in particle tracking.

## 5. Conclusion

We have described simulations and measurements of the impact of collective effects on dynamic aperture in APS. Although the accurate prediction on nonlinear beam dynamics with beam current effect is extremely challenging, the result of Elegant simulation is well agreed with experimental measurement within 5 % different. Based on the agreement of simulation result with measurement, the simulation describes current dependent nonlinear beam dynamics in detail. This result justifies the prediction on high charge injected beam dynamics in ultimate storage ring.

**Acknowledgments**
We would like to thank L. Emery (APS) and R. Nagaoka (SOLEIL) for providing helpful information and the many useful discussions. This research was supported by the U.S. Department of Energy, Office of Science, under Contract No. DE-AC02-06CH11357. This research was also supported by the Basic Science Research Program through the National Research Foundation of Korea (NRF-2015R1D1A1A01060049).

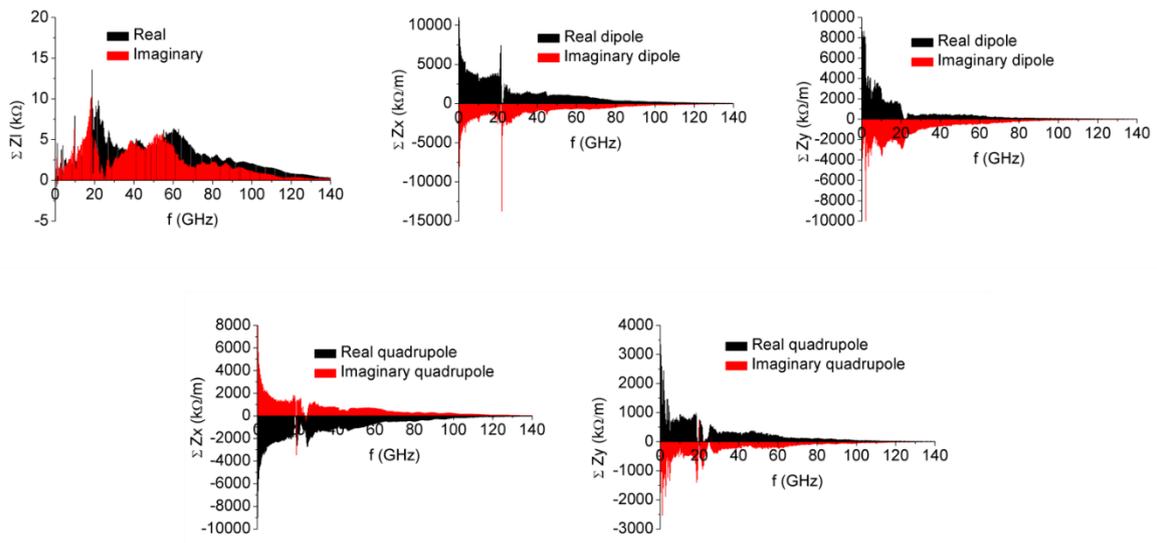

**Figure 1.** APS impedance.

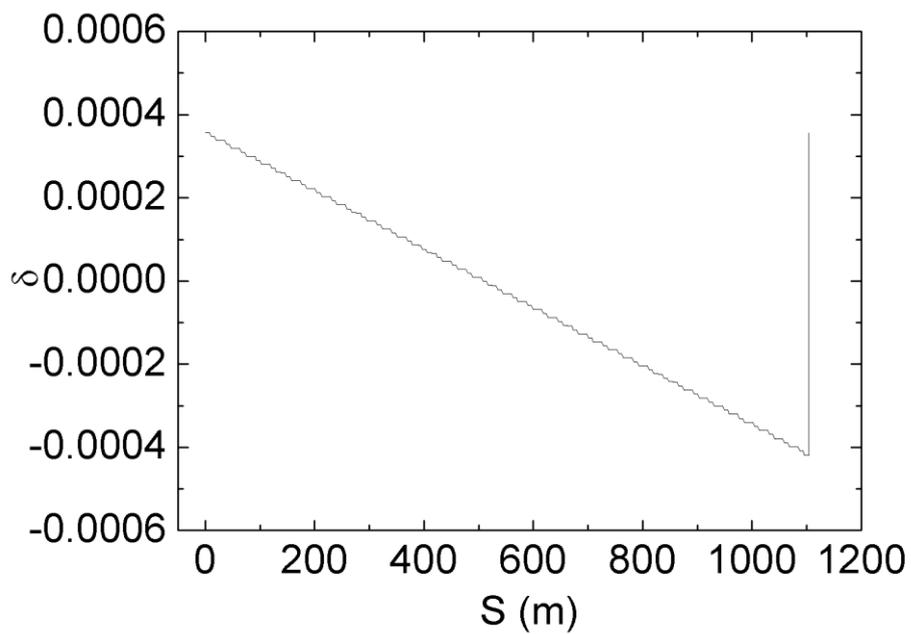

**Figure 2.** Momentum deviation along the ring.

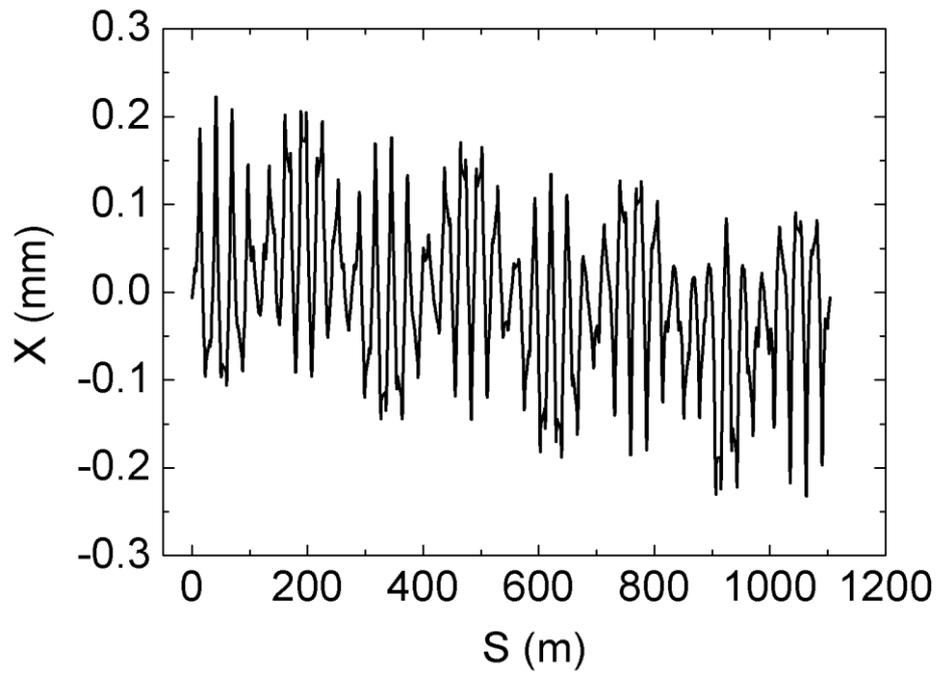

**Figure 3.** Closed orbit distortion due to energy loss and energy gain along the ring.

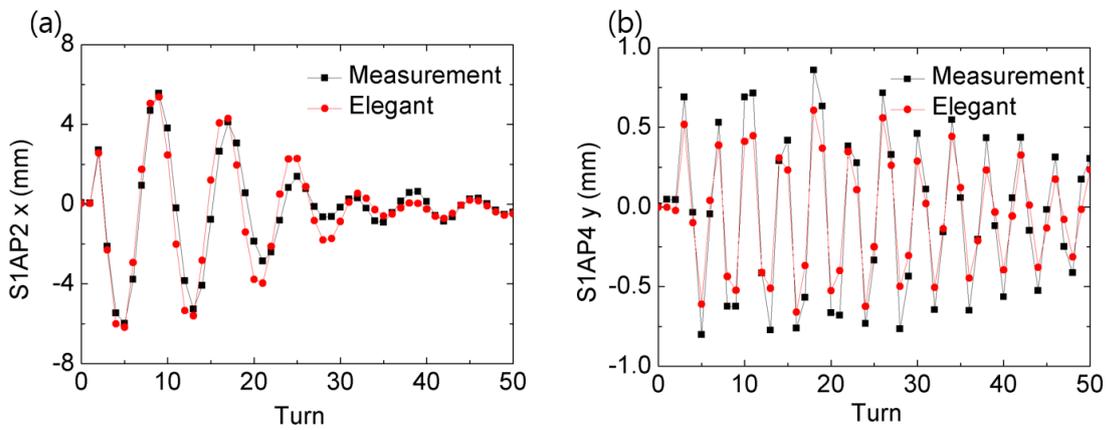

**Figure 4.** Beam oscillation by kick in horizontal and vertical planes.

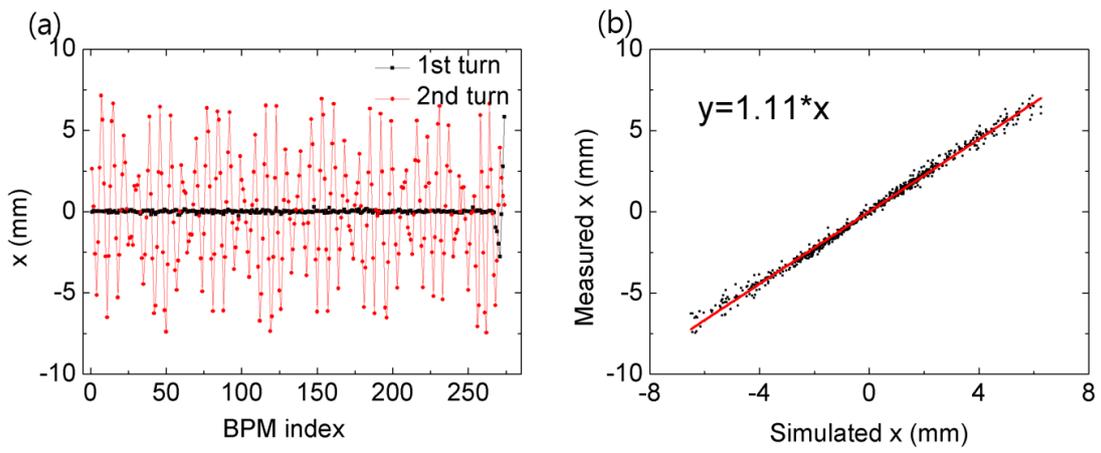

**Figure 5.** Kick strength calibration.

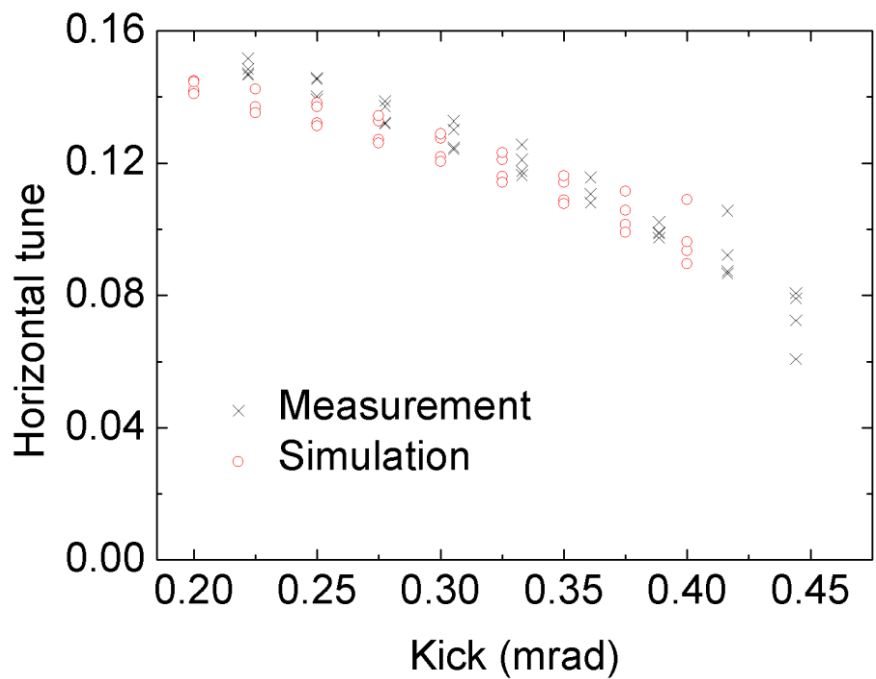

**Figure 6.** Horizontal detuning with amplitude.

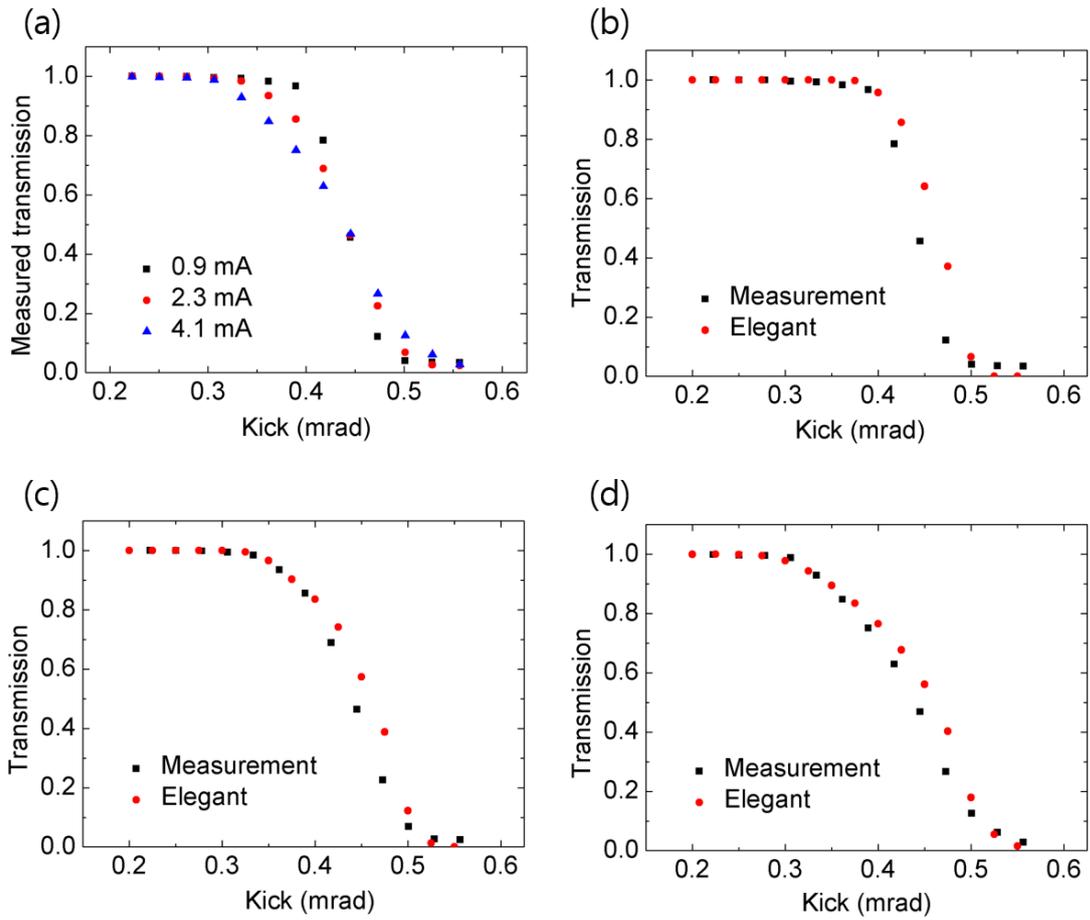

**Figure 7.** Kick aperture.

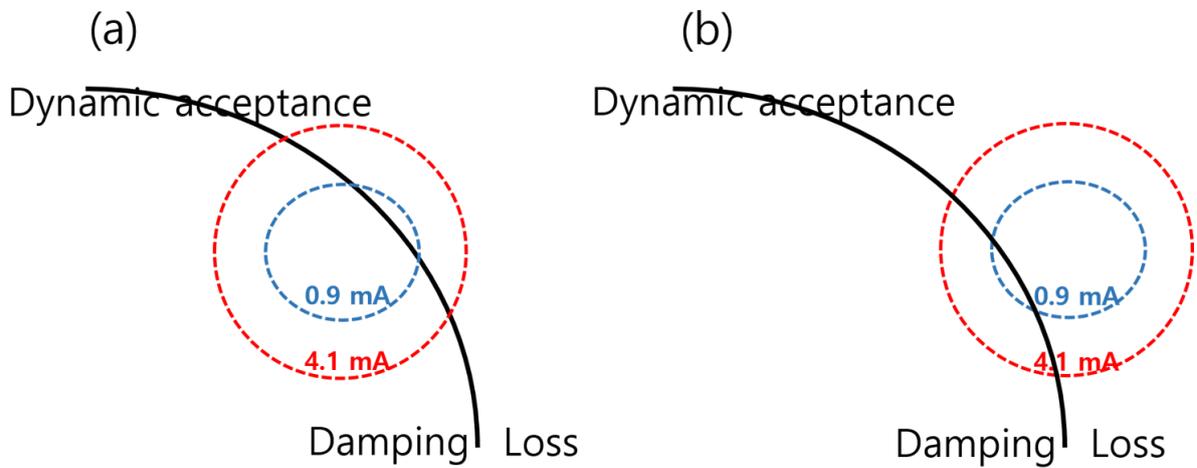

**Figure 8.** Schematics for beam loss phenomenon.

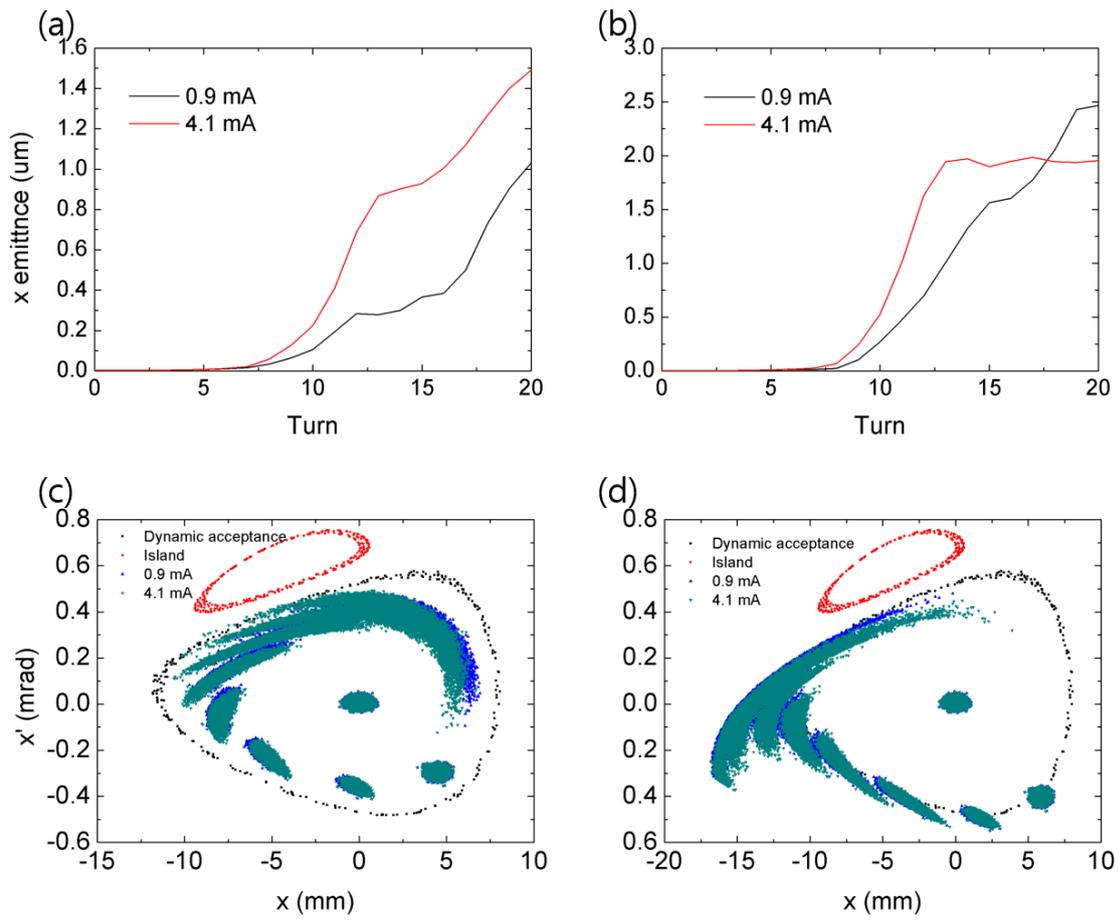

**Figure 9.** Schematics for beam loss phenomenon.

**Table 1.** Main beam parameter.

| Parameter | Value | Unit |
|---|---|---|
| Tune | 0.154 / 0.223 | |
| Coupling | 1 | % |
| Chromaticity | 6.59 / 5.85 | |
| RF voltage | 9.5 | MV |

**Table 2.** Horizontal detuning with amplitude.

| Parameter | Measurement | Simulation (Tracking) |
|---|---|---|
| $\nu_x$ | 0.154 | 0.153 |
| $\dfrac{\partial \nu_x}{\partial J_x}$ | 0.123 | 0.040 |
| $2\dfrac{\partial^2 \nu_x}{\partial J_x^2}$ | -0.670 | -0.451 |